# *Scopus'* SNIP indicator

Loet Leydesdorff
Amsterdam School of Communications Research (ASCoR), University of Amsterdam, Kloveniersburgwal 48, 1012 CX Amsterdam, The Netherlands; loet@leydesdorff.net.
Telephone: +31-20-5256598; Fax: +31-842239111.

&

Tobias Opthof
Experimental Cardiology Group, Heart Failure Research Center, Academic Medical Center AMC, Meibergdreef 9, 1105 AZ Amsterdam, The Netherlands;
t.opthof@inter.nl.net.

*Rejoinder* to Moed (2010a; at http://arxiv.org/abs/1005.4906)

1. The SNIP indicator of Elsevier's *Scopus* cannot be considered as a statistics because it is based on dividing the mean of a distribution by the value of the median of another distribution. Using this ratio, one is no longer able to indicate the standard error or to test for the significance of differences between journals or journal groups. In our opinion, one should first normalize and then one can use the mean of the resulting distribution as one statistics among others. The ISI-IF (impact factor), however, can be considered as a mean and, therefore, one would be able to use it as a statistics.

2. We did not claim that fractional counting would be "elegant and simple" in general, but that it provides a simple and elegant solution to the problem which the developers of the SNIP indicator set out to solve, namely, to control for the differences in citation behavior among fields of science (Moed, 2010b; Small & Sweeney, 1985; Zitt & Small, 2008; Zitt, 2010). Using fractional counting, distributions can be tested against each other. In Leydesdorff & Opthof (2010a) we showed this for the five journals discussed by Moed (2010b) and found, among other things, that the fractionally counted citation distributions for *Inventiones Mathematicae* and *Annals of Mathematics* were not significantly different in 2007. We are currently engaged in upscaling this indicator for developing a classification of journals based on these statistics (Leydesdorff, in preparation).

3. Moed's (2010a) claim that the SNIP indicator is valid is hollow because the indicator is based on the assumed validity of the a priori field distinctions used for the normalization. Distinguishing among fields of science on the basis of citation analysis has remained hitherto an unresolved problem, in our opinion (Leydesdorff, 2006; cf. Rosvall & Bergstrom, 2008). The ISI Subject Categories, for example, have been constructed for retrieval purposes and are not analytically based (Boyack *et al.*, 2005; Pudovkin & Garfield, 2002, at p. 1113n.; Rafols and Leydesdorff, 2009). Because the SNIP indicator is based on the assumption that the underlying field delineations are valid—*quod non*—it cannot be a valid indicator.

4. Moed (2010a) correctly noted that we could have improved the *post hoc* test by including the zeros. This is not easily possible using the *Science Citation Index*, but it may be possible using *Scopus* because this latter database contains document identifiers in both the citing and cited documents. However, we ran into several problems when using *Scopus* for the reconstruction of SNIP which made us decide to use the *Science Citation Index*.

|  | Citable items 2004-2006 | | Citing papers 2007 | | | |
|---|---|---|---|---|---|---|
|  | *SCI* | *Scopus* | SCI | Scopus before correction | *Scopus* after correction | Difference between (e) and (f) |
| (a) | (b) | (c) | (d) | (e) | (f) |  |
| *Invent Math* | 204 | 205 | 355 | 330 | 328 | 2 |
| *Mol Cell* | 923 | 922 | 8,038 | 8,428 | 8,096 | 332 |
| *J Electron Mater* | 794 | 811 | 629 | 848 | 848 | 0 |
| *Math Res Lett* | 221 | 219 | 150 | 133 | 132 | 1 |
| *Ann Math* | 165 | 172 | 512 | 434 | 433 | 1 |

**Table 1**: Citable items and citation numbers for *Scopus* and the *Science Citation Index*.

Table 1 provides a comparison of the results using the two databases for the five journals under study. In the case of *Molecular Cell*, the *Scopus* database contains more citing papers even after correction for counting only citable items. The additional set of 332 citing papers consists, among other things, of 257 "Short Surveys," 37 "Notes", and 20 "Editorials." [1] Moed wishes to correct for these citations as citations by "non-citable" items.

Let us check whether this is a good idea. We took arbitrary documents from these three categories. One "short survey", for example, is "J. Denner (2007), Transspecies transmission of retroviruses: new cases, *Virology* 369(2), 229-233." This document is classified in the *SCI* as a review. It contains 54 cited references and has been cited six times as of May 30, 2010; in addition to being published in a high-quality journal, its institutional address is a leading institute in the field (the Robert Koch Institute in Berlin).

As a test for the Notes category, we used "L.A. Amos and K. Hirose (2007), A cool look at the structural changes in kinesin motor domains, *Journal of Cell Sciences* 120(22), 3919-3927." The institutional address of this paper is the Laboratory of Molecular Biology in Cambridge (UK); the paper contains 44 references, is classified as an "article" in the *SCI*, and has been cited once. In our opinion, both these papers should be included in a citation analysis. In other words, there is good reason to question the validity of the document type distinctions in the *Scopus* database.

---

[1] Scopus lists additionally 15 letters, and 3 errata as citing *Molecular Cell* in 2007. The non-citable items for the other journals are all errata, with the exception of one editorial in the case of *Inventiones Mathematicae*.

Among the 20 editorials citing this journal, the first one was: "P.A. Jeggo and M. Löbrich (2007). DNA double-strand breaks: Their cellular and clinical impacts, *Oncogene* 26(56), 7717-7719," containing 13 references and cited 16 times since its appearance. The *Science Citation Index* also classifies this as an editorial, but the citations are nevertheless counted at the Web of Science. However, this paper is not an editorial, but an *introduction* by two leading scholars to a special review issue containing reviews about oncogene research. This paper is preceded by an editorial by one of the authors as the guest editor. Not the editorial but this introductory review provided a reference to *Molecular Cell*.

In summary, our main objection is against developing new indicators which, like some of the older ones (for example, the "crown indicator" of CWTS; cf. Opthof & Leydesdorff, 2010; Van Raan *et al*., 2010; Leydesdorff & Opthof, 2010b), do not allow for indicating error because they do not provide a statistics, but are based on dividing statistics and, in our opinion, a violation of the order of operations. Furthermore, the claim of validity is hollow because these normalizations are based on field classifications which are not valid. Both problems—(*i*) the significance of differences in impact among journals and (*ii*) field classifications on the basis of citation statistics—can perhaps be solved by using fractional counting of citations.

**References:**
Boyack, K. W., Klavans, R., & Börner, K. (2005). Mapping the Backbone of Science. *Scientometrics, 64*(3), 351-374.
Leydesdorff, L. (2006). Can Scientific Journals be Classified in Terms of Aggregated Journal-Journal Citation Relations using the Journal Citation Reports? *Journal of the American Society for Information Science & Technology, 57*(5), 601-613.
Leydesdorff, L. (in preparation). How fractional counting affects the Impact Factor: Steps towards a statistically significant classification of scholarly journals and literature.
Leydesdorff, L., & Opthof, T. (2010a). *Scopus*' Source Normalized Impact per Paper (SNIP) *versus* the Journal Impact Factor based on fractional counting of citations. *Journal of the American Society for Information Science and Technology,* in print.
Leydesdorff, L., & Opthof, T. (2010b) Normalization at the field level: Fractional counting of citations. *Journal of Informetrics* 4(4), in print.
Moed, H. F. (2010a). The Source-Normalized Impact per Paper (SNIP) is a valid and sophisticated indicator of journal citation impact, *Journal of the American Society for Information Science and Technology,* in print.
Moed, H. F. (2010b). Measuring contextual citation impact of scientific journals. *Journal of Informetrics,* in print.
Opthof, T., & Leydesdorff, L. (2010). Caveats for the journal and field normalizations in the CWTS ("Leiden") evaluations of research performance. *Journal of Informetrics* 4(3), in print.
Pudovkin, A. I., & Garfield, E. (2002). Algorithmic procedure for finding semantically related journals. *Journal of the American Society for Information Science and Technology, 53*(13), 1113-1119.

Rafols, I., & Leydesdorff, L. (2009). Content-based and Algorithmic Classifications of Journals: Perspectives on the Dynamics of Scientific Communication and Indexer Effects *Journal of the American Society for Information Science and Technology, 60*(9), 1823-1835.

Rosvall, M., & Bergstrom, C. T. (2008). Maps of random walks on complex networks reveal community structure. *Proceedings of the National Academy of Sciences, 105*(4), 1118-1123.

Small, H., & Sweeney, E. (1985). Clustering the Science Citation Index Using Co-Citations I. A Comparison of Methods, *Scientometrics 7*, 391-409.

Van Raan, A. F. J., Van Leeuwen, T. N., Visser, M. S., Van Eck, N. J., & Waltman, L. (2010). Rivals to the crown: reply to Opthof and Leydesdorff. *Journal of Informetrics, 4*(3), in print.

Zitt, M. (2010). Citing-side normalization of journal impact: A robust variant of the Audience Factor. *Journal of Informetrics*, in print.

Zitt, M., & Small, H. (2008). Modifying the journal impact factor by fractional citation weighting: The audience factor. *Journal of the American Society for Information Science and Technology, 59*(11), 1856-1860.